\documentclass{article}

 \usepackage[preprint,nonatbib]{neurips_2023}

\usepackage[utf8]{inputenc} 
\usepackage[T1]{fontenc}    
\usepackage{hyperref}       
\usepackage{url}            
\usepackage{booktabs}       
\usepackage{amsfonts}       
\usepackage{nicefrac}       
\usepackage{microtype}      
\usepackage{xcolor}         
\usepackage{graphicx}
\usepackage{float}
\usepackage{xcolor}

\title{VAE for Modified 1-Hot Generative Materials Modeling\\
{\Large A Step Towards Inverse Material Design}}

\author{%
  Khalid El-Awady\\
  Department of Computer Science\\
  Stanford University\\
  \texttt{kae@stanford.edu} \\
}

\begin{document}
\pagecolor{white}

\maketitle

\begin{abstract}
	We investigate the construction of generative models capable of encoding physical constraints that can be hard to express explicitly. For the problem of inverse material design, where one seeks to design a material with a prescribed set of properties, a significant challenge is ensuring synthetic viability of a proposed new material. We encode an implicit dataset relationships, namely that certain materials can be decomposed into other ones in the dataset, and present a VAE model capable of preserving this property in the latent space and generating new samples with the same. This is particularly useful in sequential inverse material design, an emergent research area that seeks to design a material with specific properties by sequentially adding (or removing) elements using policies trained through deep reinforcement learning. 
\end{abstract}

\section{Introduction, Problem Statement, and Dataset}

Inverse materials design, the problem of devising novel materials from first principles so they have an exact set of prescribed properties, has long been a holy grail of materials science \cite{Zunger2018}. Such a capability would allow us to tune material parameters in a way that exhibits up-to-now unrealized material behavior. The increasing performance of ML methods for computational quantum chemistry \cite{DFT2005}, along with recently available public datasets of materials and their properties \cite{MaterialsProject}, has attracted strong interest in applying AI and machine learning techniques to their analysis, and generative ones for tackling the inverse problem. 

\begin{figure}[h]
\centering
    \includegraphics[width=0.6\linewidth]{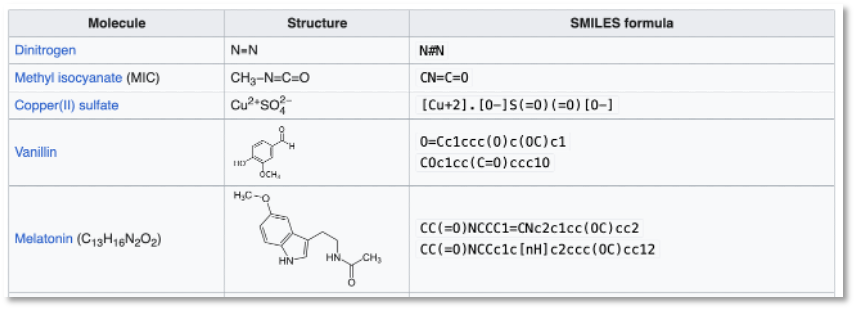}
\caption{Examples of the SMILES representation of a material. }
\label{fig:SMILES}
\end{figure}

The most common way to represent materials in AI applications has been the so-called Simplified Molecular-Input Line-Entry System (SMILES) representation \cite{Apodaca2019}. SMILES is a specification in the form of a line notation for describing the structure of chemical species using short ASCII strings \cite{WikiSMILES}. Computationally, it is a string obtained by printing the symbol nodes encountered in a depth-first tree traversal of a chemical graph. SMILES popularity stems from its ability to capture and encode information about a material including its atoms, bond structure, branching, and isotope information.  In deep learning architectures it has also been shown to work well with convolutional neural networks in capturing and encoding useful structural patterns such as ring structures. Figure (\ref{fig:SMILES}) shows some examples of the SMILES representation of materials. 

Despite its utility, SMILES has some limitations \cite{Lim2022}:
\begin{itemize}
	\item Not all materials databases include a SMILES representation for arbitrary compounds and SMILES representations are not unique. 
	
	\item It is not clear how to ensure viability of generated SMILES materials for the inverse materials design problem.
	
	\item SMILES encoding cannot easily capture relationships between compounds such as decomposition For example, consider the following material decomposition (\url{https://next-gen.materialsproject.org/materials/mp-559694?formula=Ga(MoS2)4}):
	\begin{eqnarray*}
		\mbox{Ga}(\mbox{Mo}\mbox{S}_2)_4 & = & \frac{6} {91} \cdot \mbox{Ga}_{31}\mbox{Mo}_6 + \frac{49}{55} \cdot \mbox{Mo}\mbox{S}_2 + \frac{3}{70} \cdot \mbox{Ga}\mbox{S}
	\end{eqnarray*}

\end{itemize}

\begin{figure}[h]
\centering
    \includegraphics[width=0.8\linewidth]{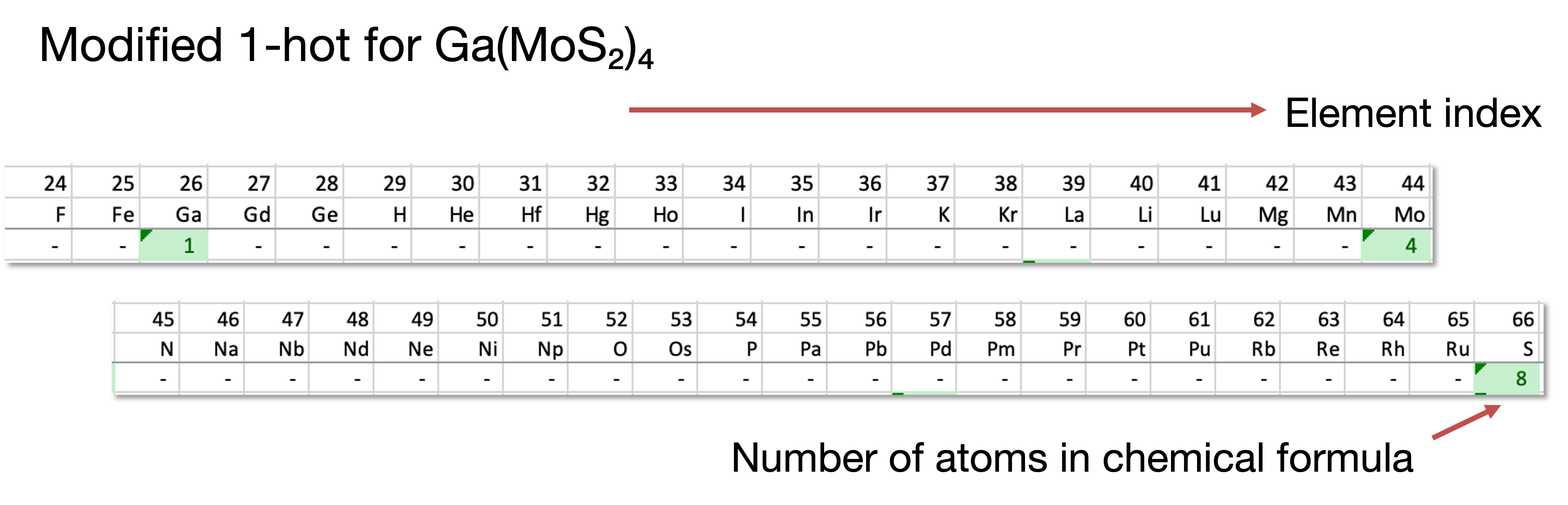}
\caption{Schematic representation of the modified 1-hot vector representation for $\mbox{Ga}(\mbox{Mo}\mbox{S}_2)_4$.}
\label{fig:modonehot}
\end{figure}

The decomposition property (and its inverse property - synthesis) is a key one of interest to us. Recently, a number of researchers (including the author) have demonstrated the use of reinforcement learning to discover policies to synthesize viable materials using sequential additions or deletions of elements \cite{Pan2022, Lacomb2023}. A more natural material representation that readily captures decomposition is the modified 1-hot material encoding. In our dataset there are 89 unique elements comprising all the materials. We sort the elements alphabetically by symbol and assign each a sequential index (Ac = 1, Ag = 2, Al = 3, ...). We then represent each material as a 89-length vector where entry $j$ is the number of atoms of that element in the molecule. For example the material, $\mbox{Ga}(\mbox{Mo}\mbox{S}_2)_4$ would have a 1 in position 26 (Gallium), a 4 in position 44 (Molybdenum), an 8 in position 66 (Sulfur), and zeros in all remaining entries in the vector. This is shown schematically in figure (\ref{fig:modonehot}). 

\begin{figure}[h]
\centering
    \includegraphics[width=0.6\linewidth]{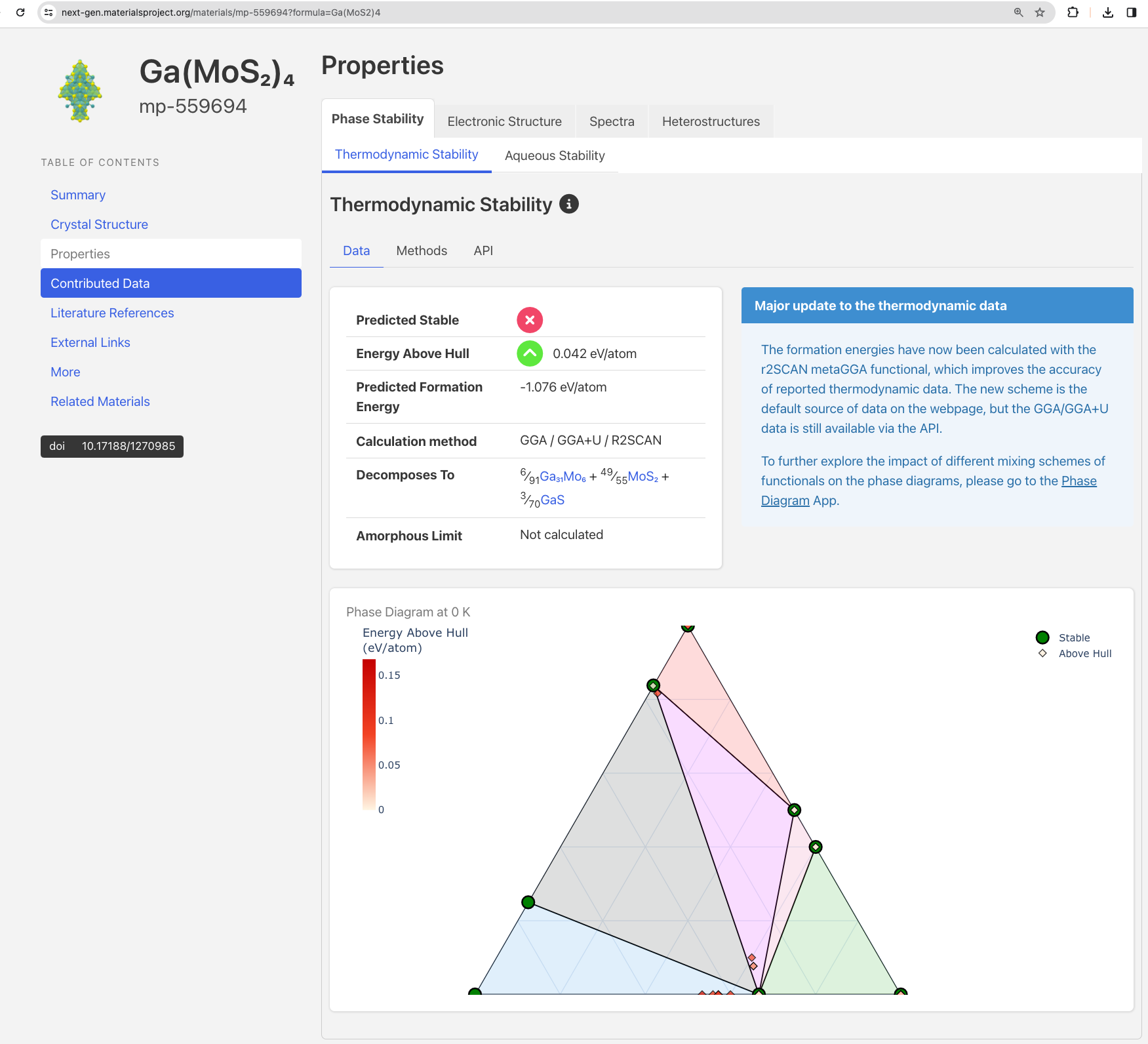}
\caption{Example material entry in the Materials Project database: $\mbox{Ga}(\mbox{Mo}\mbox{S}_2)_4$.}
\label{fig:MPentry}
\end{figure}

In this work we explore a generative model for a modified 1-hot representation of materials. We show that a variational autoencoder (VAE) architecture is capable of capturing the decomposition property and preserving it in the latent space. For our dataset we will use the Materials Project database \cite{MaterialsProject}, an open-access database of approximately 154,000 unique materials relevant to battery research. Each entry contains a variety of properties pertaining to thermodynamic stability, electronic structure, magnetic properties, and others. The dataset also includes the constituent components in the dataset, if the material decomposes into others, and their relative weighting in the combination. An illustrative example of one material,  $\mbox{Ga}(\mbox{Mo}\mbox{S}_2)_4$, is shown in figure (\ref{fig:MPentry}). 

\section{Related Work}

The prior work of Gomez, et. al., \cite{Gomez2018} is usually cited as the first demonstration of the use of VAEs in new molecule generation. They used a SMILES representation to produce a latent representation of a material and show that the latent variables correlate well with certain material properties and therefore have efficacy in prediction. The work, though, suffers from the fact that from a generative perspective there is no enforcement that a sampled SMILES representation is either viable or valid. 

GAN-based models were introduced by Nouria, et. al. \cite{Nouria2018}. In their work the authors propose a GAN arcihtecture they call, CrystalGAN, whose  goal is to generate novel ternary metal hybrids from their observed binary structures. The authors demonstrate an ability to generate these ternary crystals, and though it has been scientifically known that novel ternary compounds can be predicted from known
binaries, whether this approach can be applied to predict complex crystal structures is uncertain. Therefore ensuring general viability is still an open question. A further review of GAN application in inverse material design can be found in \cite{Jabbar2022}. 

While both the VAE and GAN approaches have shown promise, challenges remain including \cite{Lu2022, Menon2022}: 
\begin{itemize}
	\item generating materials with more than a small handful of target properties, 
	\item constraining sampling of the models to synthetically viable materials, and 
	\item dealing with what are still relatively small datasets (hundreds of thousands of data samples, not billions).
\end{itemize}

Our work deals primarily with addressing the viability question, specifically by exploring generative models that preserve material decomposition which can be exploited in the subsequent synthesis of viable new materials. 

\section{Approach}
\begin{figure}[h]
\centering
    \includegraphics[width=1.1\linewidth]{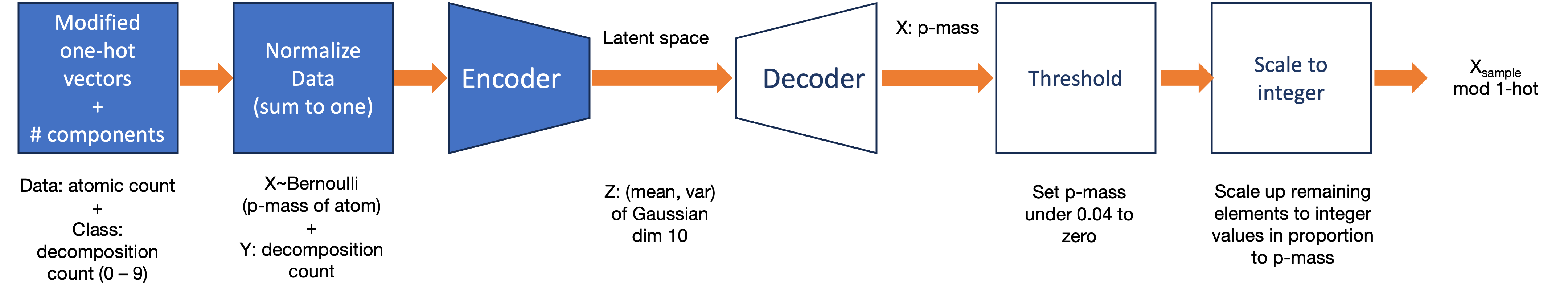}
\caption{Model architecture.}
\label{fig:modelArchitecture}
\end{figure}

In the interest of starting simply, we choose a VAE-based architecture. To formulate a probabilistic model for the material generation process we examine the dataset and notice the following:
\begin{itemize}
	\item Materials' modified 1-hot vectors are sparse. The most complex material in the dataset has 10 unique elements, though 96\% of the materials have 6 or fewer elements.
	
	\item Chemical formulae are invariant under scaling and generally materials are represented such that the atomic counts of different elements have a greatest common divisor of 1. 
	
	\item Examination of the element pair-wise correlation matrix across the dataset shows that here is almost no correlation between the presence of elements in a given material. Of the 3,916 unique off-diagonal element pairs only 7 have correlations exceeding 0.2 in magnitude. The largest is the pair (Nb, O) with a correlation coefficient of 0.54. But for practical purposes we model the elements as uncorrelated.
	
\end{itemize}
Accordingly we postulate a model of our material as a set of independent Bernoulli random variables indicating the presence of the atom in the material or not. To account for the relative weight of the atom we normalize the entries in the modified 1-hot vector so that it always sums to one in the dataset. The full model architecture is shown in figure (\ref{fig:modelArchitecture}).

A summary of the encoding / sampling procedure is as follows:
\begin{itemize}

	\item {\bf Material Representation:} the chemical formula of the material is converted to a modified 1-hot representation. For example the material, $\mbox{Ga}(\mbox{Mo}\mbox{S}_2)_4$ would have a 1 in position 26 (Gallium), a 4 in position 44 (Molybdenum), an 8 in position 66 (Sulfur), and zeros in all remaining entries in the vector. 
	
	\item {\bf Normalization:} the modified 1-hot vector is normalized to sum to one. For example the material, $\mbox{Ga}(\mbox{Mo}\mbox{S}_2)_4$ which has a total of 13 atoms is normalized by dividing the elements of the modified 1-hot vector by 13. This normalized vector will be our data samples, $x^{(i)}$. We also include a class variable, $Y$, representing the number of decomposition components of the given material data point. In our dataset materials decompose into between 0 and 9 components so that there are 10 total classes.
	
	\item {\bf Encoder:} we propose a deep network architecture similar to the one developed in class for MNIST digit generation. This was selected due to the similarity of our normalized 1-hot representation to a gray-scale representation of images. The architecture comprises the following layers: 
	\begin{eqnarray*}
		\mbox{Linear} \, \rightarrow \, \mbox{ELU} \, \rightarrow \, \mbox{Linear} \, \rightarrow \, \mbox{ELU} \, \rightarrow \, \mbox{Linear} 
	\end{eqnarray*}
	The output of the last layer is a $2 \times n$ vector representing the mean and variance, $(\mu_\theta(x), v_\theta(x))$ of a latent variable, $Z \sim \mathcal{N}(\mu_\theta, v_\theta)$ of dimension, $n$. The classifier has a similar architecture to the encoder with a single output representing the material's class (number of components). The main parameter we tune is the number of nodes per hidden layer which we keep common between all layers and for both encoder, decoder, and classifier for simplicity. Tuning of this parameter is discussed in the section on results. 
	
	The training of the VAE utilizes a negative ELBO loss function. 
	
	\item {\bf Decoder:} The decoder is a mirror image of the encoder, comprising the following layers:
	\begin{eqnarray*}
		\mbox{Linear} \, \rightarrow \, \mbox{ELU} \, \rightarrow \, \mbox{Linear} \, \rightarrow \, \mbox{ELU} \, \rightarrow \, \mbox{Linear} 
	\end{eqnarray*}
	
	\item {\bf Threshold}: In the generation process, the decoder returns a continuous vector whose elements  are interpreted as the contribution of that element to the overall atomic count. To convert this to a vector with a discrete number of unique atoms we apply a threshold, $T$,  below which elements are zeroed out. The value of the threshold is tuned to match the distribution of the number of elements per material in the dataset. We find a value of 0.04 works well. The analysis for this parameter is discussed in more detail in the results section.
	
	\item {\bf Scaling:} after thresholding, the sample is converted to a material representation by scaling up all entries by $1/T$, then computing the greatest common divisor, $g$, for the elements and reducing the formula to its simplest form. 

\end{itemize}

The following section includes a description of the model tuning process and analysis of the results.

\section{Tuning and Results}

\subsection{Model Tuning}

We use a latent variable dimension of 10 and train for 2,000 epochs. Due to the sparsity of materials with 8 or 9 elements we oversample them in the dataset. After training the model we collect 10,000 samples from the model and compare some statistical properties of the two datasets. To assess the quality of the model we consider three metrics:
\begin{enumerate}

	\item Negative ELBO: the loss function used in the model training assessed on the training set which comprises 20\% of the samples in our dataset. 
	
	\item KL divergence between the probability distribution of elementwise prevalence of elements in the dataset vs the samples. The distribution of elementwise prevalence is computed by summing the modified 1-hot vectors across samples to obtain a total count of element atoms in the dataset then normalizing across all elements. 
	
	\item Comparison of the distribution of number of elements in a sampled material vs the dataset.

\end{enumerate} 

\begin{figure}[h]
\centering
    \includegraphics[width=0.7\linewidth]{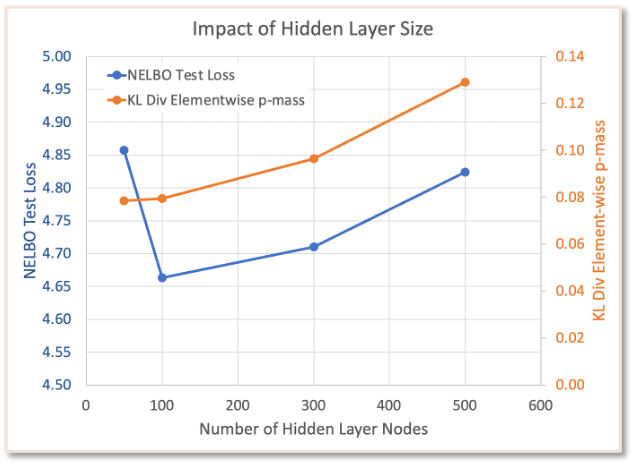}
\caption{Impact of the size of the hidden layer in the VAE on the model performance. We choose 100 hidden nodes as the 'optimal' value.}
\label{fig:hiddenLayer}
\end{figure}

Figure (\ref{fig:hiddenLayer}) shows the impact of the size of the hidden layer on the quality of samples.  We vary the number of nodes in the hidden layer from 50 to 500 and compare the test set NELBO and KL divergence of elementwise prevalence  between the model and data. We see that a hidden layer of 100 nodes provides a good combination of the two metrics and that is the chosen value for subsequent analysis. Further, this and subsequent plots were also used to tune the threshold parameter to $T = 0.04$.  

\begin{figure}[h]
\centering
    \includegraphics[width=0.7\linewidth]{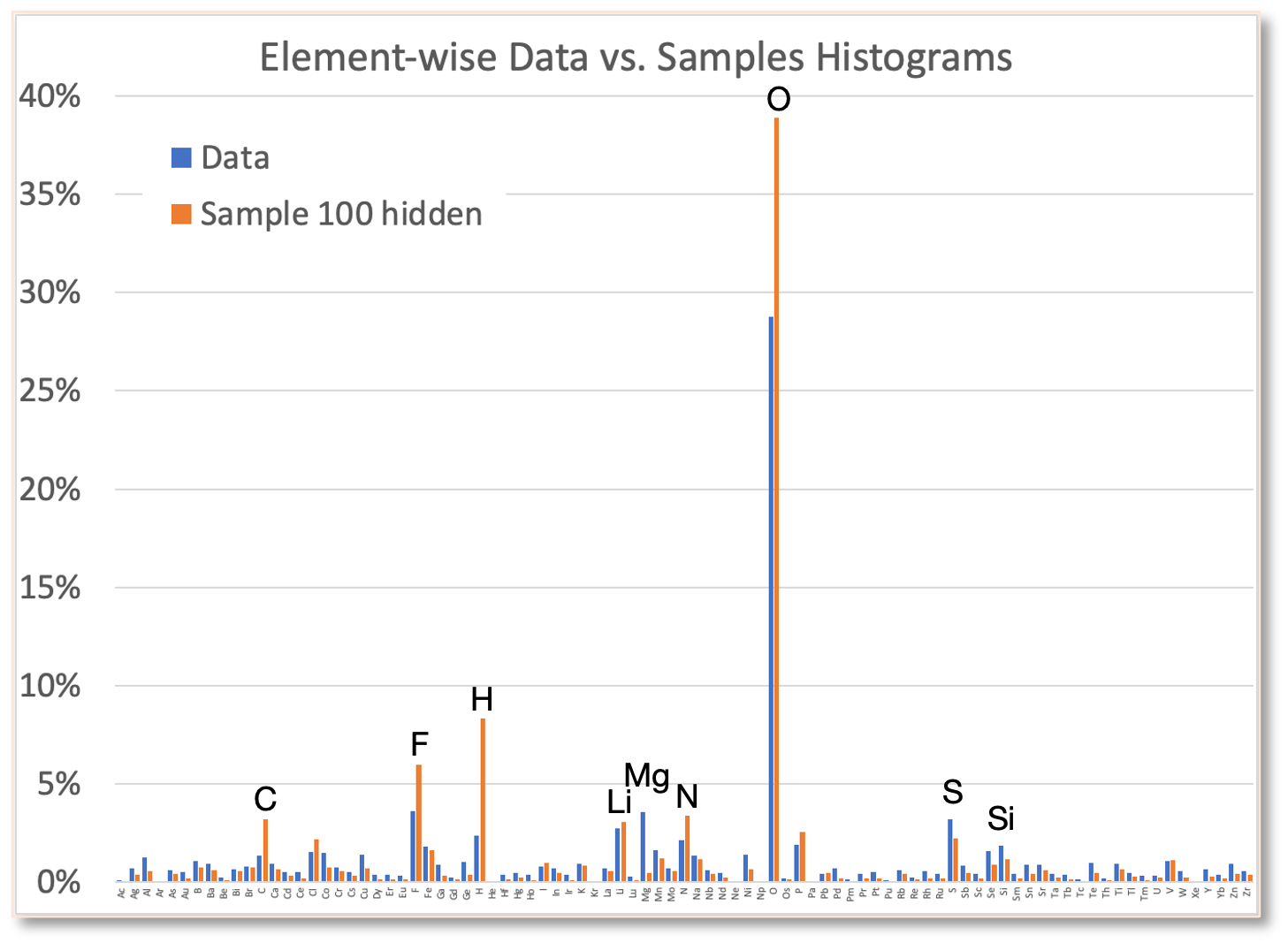}
\caption{Comparison of the elementwise prevalence probability distribution between the dataset and model. Overall the KL divergence is low (0.08).}
\label{fig:elementwise}
\end{figure}

Figure (\ref{fig:elementwise}) shows a comparison of the elementwise prevalence probability vectors between the data and samples. The KL divergence between the two is small: 0.08. But we do notice some over-representation on certain elements such as O, F, C, N, Li, and P. And certain other elements are slightly under- represented including Mg, Co, Cu, S, and Si. 

\begin{figure}[h]
\centering
    \includegraphics[width=0.7\linewidth]{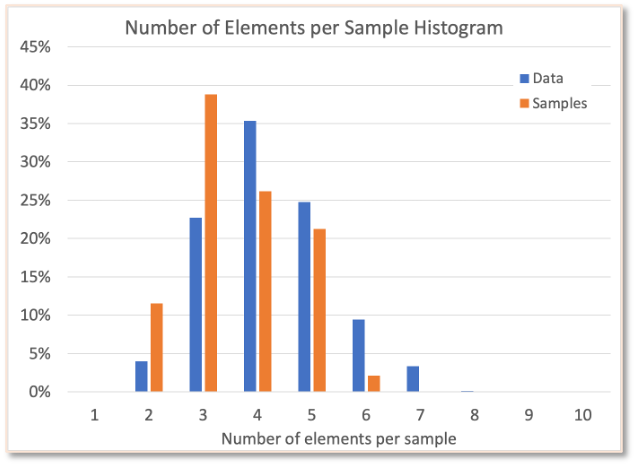}
\caption{Comparison of the distribution of the number of elements in a material in the dataset and generated samples.}
\label{fig:numElements}
\end{figure}

Figure (\ref{fig:numElements}) shows the distribution of the number of elements in a material in the data and model samples. We see a good match with the exception of some under-representation in the model of materials with a high number of elements (over 6). We are comfortable with this discrepancy since viability and synthesizability concerns generally grow quickly with complex many-atomed materials. 

\subsection{Analysis of Decomposition in the Latent Space}
With a tuned model we proceed to analyze how well the model captures and preserves decomposition. Our dataset  includes the constituent datapoints making up any given material along with their weights. Namely for any datapoint, $x^{(i)}$, we know
\begin{eqnarray*}
	x^{(i)} & = & \sum_{k=0}^{c_i} a_k x^{(k)} 
\end{eqnarray*}
where $c_i$ is the number of components making up material, $x^{(i)}$, and that varies from 0 to 9 in our dataset, and $a_k$ are known constants. 

\begin{figure}[h]
\centering
    \includegraphics[width=1.0\linewidth]{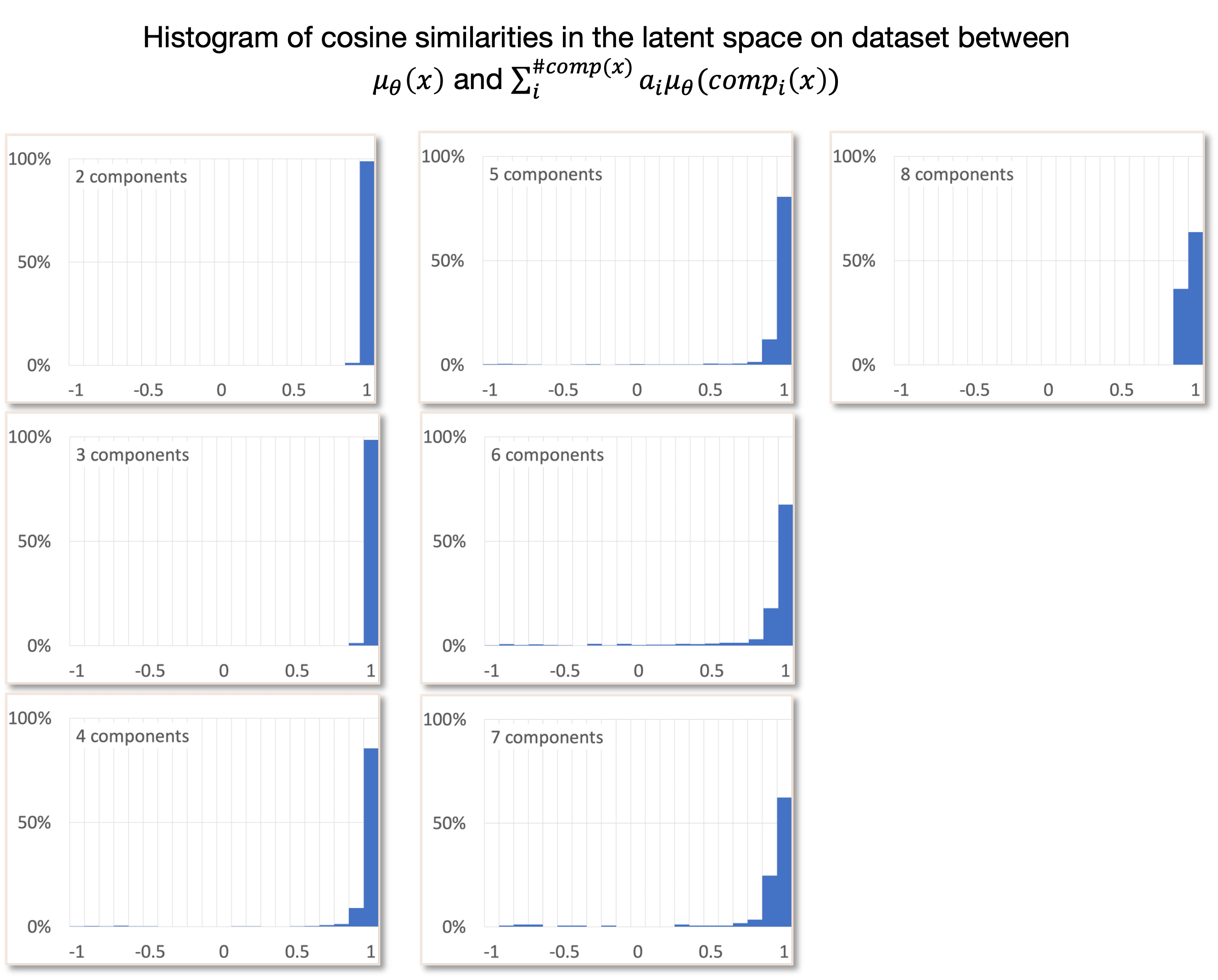}
\caption{Latent space decomposition similarity of means and their weighted combined constituents.}
\label{fig:similarity}
\end{figure}

For each of the dataset samples we utilize our encoder to compute the latent vector means: $\mu_\theta(x^{(i)})$. Next we investigate how those vectors compare to the re-constituted latent vector composed of the weighted sum of the latent mean components. For comparison we use the cosine similarity. Thus for each material that decomposes into 2 or more components we compute:
\begin{eqnarray*}
	\mbox{latent space decomposition similarity} & = & < \mu_\theta(x^{(i)}),  \sum_{k=0}^{c_i} a_k \mu_\theta(x^{(k)}) >
\end{eqnarray*}
We plot the histogram of similarities by number of components for the dataset. The results are shown in figure (\ref{fig:similarity}). 

We see remarkable similarity preservation of the decomposition property in the latent space. This indicates that our latent representation can learn and preserve the decomposition property and linear manipulations on latent vectors are largely reflective of similar manipulations on the original material.

\section{Conclusions}
In this study we present evidence in support of using modified 1-hot vector representations of materials versus the more commonly used SMILES representation. Such representations conveniently preserve decomposition, and by extension synthesis. This is particularly useful in sequential inverse material design, an emergent research area that seeks to design a material with specific properties by sequentially adding (or removing) elements using policies trained through deep reinforcement learning. We anticipate that RL policies designed to manipulate our latent space will provide advantages in terms of flexibility of the action space on the continuous latent space over directly trying to manipulate the modified 1-hot vectors while trying to simultaneously preserve constraints on material viability. 

The next steps for this work could include:
\begin{itemize}

	\item Investigation of the predictive power of the latent space on other material properties of interest, such as the phase diagram. 

	\item  Augmention of the dataset with other material properties and analysis of the ability of the latent space to capture those as well. Would the latent space be able to capture non-linear effects of material property superposition? 

	\item Exploration of training RL policies on the latent space and analyzing whether these policies reduce the complexity of viability constraints on inverse material design. 

\end{itemize}

\medskip
{
\small

\bibliographystyle{plain}

}


\end{document}